\documentclass[twocolumn,showpacs,preprintnumbers,amsmath,amssymb]{revtex4}
\usepackage{graphicx}

\begin{document}
\title{Toward protocols for quantum-ensured privacy and secure voting}
\author{Marianna  Bonanome$^{1}$, Vladim\'{\i}r Bu\v{z}ek$^{2,3}$, Mark Hillery$^{4}$ and M\'{a}rio Ziman$^{2,3}$ }
\affiliation{
$^{1}$Department of Applied Mathematics and Computer Science, New York City College of Techology, 300 Jay Street,  Brooklyn, NY 11201 USA \\
$^{2}$Research Center for Quantum Information, Slovak Academy of Sciences,
       D\'{u}bravsk\'{a} cesta 9, 845 11 Bratislava, Slovakia\\
$^{3}${\em Quniverse}, L{\'\i}\v{s}\v{c}ie \'{u}dolie 116, 841 04 Bratislava, Slovakia\\
$^{4}$Department of Physics, Hunter College of CUNY, 695 Park Avenue,  New York, NY 10021 USA
 }

\begin{abstract}
We present a number of schemes that use quantum mechanics to preserve
privacy, in particular, we show that entangled quantum states can be useful in maintaining privacy.
We further develop our original proposal [see Phys. Lett. A {\bf 349}, 75 (2006)] for protecting
privacy in voting, and examine
its security under certain types of attacks, in particular dishonest voters and external eavesdroppers.
A variation of these quantum-based schemes can be used for multi-party
function evaluation. We consider functions corresponding to
group multiplication of $N$ group elements, with each element chosen by a different party.
We show how quantum mechanics can be useful in maintaining the privacy of the choices
group elements.
\end{abstract}

\maketitle

\section{Introduction}
There are many situations in which maintaining the privacy of
information is important.  One example is voting; a voter (let us
call him Vincent) does not want either other voters or the person
counting the votes to know how he voted.  Another possible
situation is one in which a number of parties want to pool their
financial resources to purchase, perhaps, a company.  They need to
find out if the total amount of money they have is sufficient, but
each individual does not want the others to know how much he or
she has.

Quantum mechanics has proven to be a useful basis of novel communication schemes.  In
particular, quantum key distribution uses the laws of physics as the basis for a scheme to
distribute secure cryptographic keys \cite{gisin}.
Here we would like to discuss whether quantum mechanics can be used to protect
privacy as well. In particular, we shall examine the role quantum mechanics can play
in voting schemes and in a special form of distributed function computation.
The elementary primitives for privacy are the anonymous broadcast channels.
An anonymous one-to-many broadcast channel is one in which each of the parties
can send a message to all of the others, but only the person who sent the message
will know who sent it, i.e. his identity remains hidden to each receiver.
One solution \cite{chaum,bos} is based on DC-nets, which solves the so-called dinning
cryptographer's problem (originally formulated by D.Chaum in Ref.~\cite{chaum}),
provided that the communication is secured by a one-time pad.
As discussed in Refs.~\cite{christandl,bouda,brassard}
the quantum-based anonymous broadcast of classical information does not provide
us with additional security beyond that provided by classical protocols.
However, it is possible to anonymously broadcast quantum information,
in particular, as is shown in Ref.~\cite{christandl,bouda}
an unknown state of a quantum system (i.e. quantum information) can be
teleported anonymously, so that the identity of the sender of the quantum
information remains hidden.

The paper is structured as follows: In  Sec.~II we review the
quantum-based voting protocols.  In this section and the following one, we assume that everyone
participating in the protocol is honest but curious, i.e. they follow the steps of the protocol, but if
any extra information comes their way, they will have a look.  In Sec.~III, we show how voting is a
special case of a kind of distributed function evaluation.  In Sec.~IV, we change the adversary model
and look first at the case of dishonest voters, and then at the issue of a eavesdropper who wishes to
learn how one of the voters voted.  We summarize our results
in Sec.~V. A detailed analysis of an attack by a cheating voter can be found in Appendix.

\section{Anonymous voting}
Let us assume that there are $N$ parties, and they are each to vote ``yes'' or``no'' on
some question.  Besides the voters, there is also an authority (let us call her Alice)
who provides the resources for voting and counts the votes.  Throughout the paper, we shall assume
that the authority is honest but curious, that is the authority will follow the protocol, but if any information
is available to her, she will have a look.
Some desirable features that we might want our voting procedure
to satisfy are (for details see Ref.~\cite{schneier}):
\begin{enumerate}
\item Privacy - only the individual voter should know how he or
she voted \item Security - each voter can vote only once and
cannot change someone else's vote \item Verifiability - each voter
can make sure that his or her vote has been counted properly, but
simultaneously cannot prove to anyone else how he was voting \item
Eligibility - only eligible voters can vote.
\end{enumerate}
We shall mainly be discussing the first requirement, but we shall suggest a method of
guaranteeing the second requirement as well. The analysis of the other two conditions
is beyond the scope of the present paper.
 A considerable effort in classical cryptography has gone
into designing voting systems, but here we shall only consider quantum based approaches.
It is important to say that the above list of conditions can be extended and there are
different variations of properties the voting should satisfy. Depending on the specified
conditions there exists unconditionally secure classical protocols, but their description
is beyond the scope of this paper.

There have been two quantum-based voting schemes proposed independently \cite{vaccaro,hillery}.
The quantum-based voting scheme proposed by J.\ Vaccaro, J.\ Spring, and A.\ Chefles \cite{vaccaro}
makes use of multiparty states whose total particle number is definite, but the total number of
particles possessed by an individual voter is not fixed.  The votes are encoded
in a phase.  We shall discuss here the schemes originally proposed in our earlier paper \cite{hillery}, one
of which also encodes votes in a phase, but in this case each voter has a fixed number of
particles.  In what follows we will study in detail two types of voting schemes: A traveling ballot scheme, and a
distributed ballot scheme.  As was mentioned in the Introduction, in this section we shall assume
that everyone is honest but curious, and we will focus on the privacy condition.

\subsection{Travelling ballot}
Let us first consider the traveling ballot scheme. We shall consider $N$ voters
(Vincent.1, Vincent.2 $\dots$, Vincent.N) and an authority to count the votes.
The authority (Alice) begins by preparing the entangled two-qudit state ($D>N$)
\begin{equation}
|\Psi\rangle =\frac{1}{\sqrt{D}}\sum_{j=0}^{D-1}|j\rangle_{a} |j\rangle_{b} .
\label{ballot_travel}
\end{equation}
The authority holds the first qudit and sends the second one to Vincent.1. He now performs one
of two operations:  if he wants to vote ``yes'', he performs the operation $E_{+}$,
where $E_{+}|j\rangle = |j+1\rangle$ (the addition is modulo $D$), and if he wants to vote
``no'' he does nothing (the identity operator).  Vincent.1 then passes the qudit on to Vincent.2, who makes the same choice
and sends it further. Finally, Vincent.N sends the qudit back to Alice (the authority).
The authority's final two-qudit state is
\begin{equation}
|\Psi_{m}\rangle =\frac{1}{\sqrt{D}}\sum_{j=0}^{D-1}|j\rangle_{a} |j+m\rangle_{b} ,
\end{equation}
where $m$ is the number of ``yes'' votes.  We have that $\langle\Psi_{m}|\Psi_{m^{\prime}}
\rangle = \delta_{m,m^{\prime}}$, so that if Alice measures the final state
in the basis $\{ |\Psi_{m}\rangle |m=0,\dots D-1 \}$, she will be able to determine
the number of ``yes'' votes. Let us note a number of things about this scheme.

\noindent\newline
1) The privacy is guaranteed by the fact that
there is ``no'' information in the state $|\Psi_{m}\rangle$ about who voted ``yes'' and who voted ``no''.
In addition, during the entire time the second qudit is traveling (before it is returned to the
authority), the reduced density matrices of all voters and the authority is $\rho = (1/D) I$,
where $I$ is the identity matrix.  That means that during the voting process, neither the
voters nor the authority can determine how the voting is progressing.  In particular, Vincent.2
cannot determine by examining the particle he receives from Vincent.1 how he voted.  Therefore,
the scheme maintains the privacy of the voting process.

\noindent\newline 2) This is {\em stronger} security than that
provided by a naive classical scheme.  In that scheme, a ballot
goes from voter to voter, and each voter enters into it his vote, $0$
for``no''and $1$ for ``yes'', plus a random number. At the end the
ballot goes back to the authority, and everyone sends their random
number to the authority, who then subtracts their sum from the
total number on the ballot to arrive at the number of ``yes''
votes. If the random numbers remain secret, the scheme insures
privacy, but if the random number of one of the voters, Vincent.2,
for example, becomes known,  then the voter who voted just before,
i.e. Vincent.1, and the one who voted just after him (Vincent.3) can determine
Vincent.2's vote.  The quantum scheme does not require the use of
secret information, which can become compromised.

A traveling ballot can also be used for, what was called in Ref.~\cite{vaccaro}, an anonymous
survey.  This can be used to compute the average salary of a group of people without
learning the salary of any individual.  One uses a traveling ballot, and each person ``votes'' a
number of times that is proportional to their salary, e.g. one vote means $10,000$ Euros, two
means $20,000$ Euros, etc.  The authority counts the number of votes and divides by the
number of voters to find the average, but the information about individual salaries is
available neither to the authority nor to the individual voters.

\subsection{Distributed ballot}
For the case of a distributed ballot the framework is the same, i.e. we shall suppose
that there are $N$ voters and an authority, who counts the votes.
The authority prepares an entangled $N$-qudit ballot state \cite{hillery}
\begin{equation}
\label{ballotstate}
|\Psi\rangle = \frac{1}{\sqrt{D}}\sum_{j=0}^{D-1}|j\rangle^{\otimes N} ,
\end{equation}
where the states $\{ |j\rangle\ |j=0,\ldots D-1\}$ form an orthonormal basis
for the $D$-dimensional space of an individual qudit, and $D>N$.
A single qudit is now distributed to each of the voters.
In order to vote``no''a voter does nothing, and to vote ``yes'',
he applies the operator
\begin{equation}
\label{yesop}
F=\sum_{k=0}^{D-1}e^{2\pi ik/D} |k\rangle\langle k| ,
\end{equation}
to his qudit.  Note that at all times during the voting procedure the reduced density matrix of
the qudit of a single voter is $\rho =(1/D) I$, so that he can infer nothing about
the votes of the other voters.  All of the qudits are then sent back to the authority, whose state
is now (if $m$ people voted ``yes'')
\begin{equation}
|\Psi_{m}\rangle = \frac{1}{\sqrt{D}}\sum_{j=0}^{D-1} e^{2\pi ijm/D} |j\rangle^{\otimes N} .
\end{equation}
The states $|\Psi_{m}\rangle$ are orthogonal for different values of $m$ and hence can be
perfectly distinguished.  Consequently, the authority can determine the number of ``yes'' votes.
Note that the states $|\Psi_{m}\rangle$ contain no information about who voted ``yes''; they
encode only the total number of ``yes'' votes.  Again, voter privacy is protected. 

An interesting variant on this procedure was proposed by Dolev, \emph{et al}. \cite{dolev}.  In
their scheme, the ballot state is locally unitarily equivalent to the state in
Eq.~(\ref{ballotstate}), the ``yes'' vote is described by operation $E_+$
($E_{+}|j\rangle = |j+1\rangle$) and $D=N+1$. In particular, the ballot state
is
\begin{equation}
\label{modDballot}
|\Phi\rangle = \frac{1}{\sqrt{D^{N-1}}}
\sum_{l_1+\dots+l_N=0\ {\rm mod}D}
|l_1\rangle\dots|l_N\rangle .
\end{equation}
We define $b_{n}=0$ if the $n^{\rm th}$ voter voted``no''and $b_{n}=1$ if the $n^{\rm th}$ voter
voted ``yes'', then the state after the voting is
\begin{equation}
|\Phi^{\prime}\rangle = \frac{1}{\sqrt{D^{N-1}}}
\sum_{l_{1}+\ldots l_{N}=0\ {\rm mod}D}
|l_{1}+b_{1}\rangle \ldots |l_{N}+b_{N}\rangle ,
\end{equation}
where the addition inside the kets is mod $D$.  Each voter now measures his qudit in the
computational basis getting the outcome $x_j=l_j+b_j$ containing his vote ($b_j$) and
a random number ($l_j$) added to it, but these numbers have the property that they add to zero
mod $D$, i.e. $\sum_j l_j = {\rm mod} D$.
Each voter announces the result of his  measurement and the total sum
$x=\sum_{j} (l_j+b_j)=\sum_j b_j$ gives the result of the voting.
That is, each voter adds all of the announced results mod $D$,
and the result $x$ is the total number of ``yes'' votes.

This scheme can be modified to perform one-to-many anonymous
broadcast, sending $\log D$ bits of information. Consider $N$ parties sharing
the state $|\Phi\rangle$, and let the sender performs the operation $E_{+}^{m}$.
This will result in the new state in which the number $l_{1}, l{2}, \ldots l_{N}$ sum
to $m$ mod$D$. Measuring in the computational basis and publicly announcing the results 
will enable each of party to reconstruct the message $m$. In a sense this protocol
provides a quantum solution to dining cryptographer's problem \cite{chaum}.

\section{Distributed group multiplication}
Maintaining privacy on decision making (e.g., voting) can be considered as a part of a more general task
- multi-party function evaluation.  In particular, voting and an anonymous survey can be viewed as each
participant picking a member of a cyclic group with the object being to compute the sum of all of the 
chosen members and doing so in such a way that the participants' choices are not revealed.  We would
like to show that a similar procedure works for computing the product of group elements chosen by the
participants for any group.  That is, voting is just a special case of distributed group multiplication.
Throughout this section we shall assume that everyone is honest but curious.

This problem is related to that of secure function evaluation.  Suppose that Donna has a 
device that will evaluate the function $f(x,y,z)$.
Alice, who has the input $x$, Bob, who has the input $y$, and Charlie, who has input $z$,
would like to know the value of $f(x,y,z)$, but each of them wants Donna and the other two
participants to know as little about their input as possible.  In fact, ideally Donna would
know only as much as she can infer from knowing the value of $f(x,y,z)$, and each of the
other parties would only know as much as they could infer from knowing $f(x,y,z)$ and
their own input.  Note that voting is a special case of this problem in which the variables
take only the values $0$ and $1$, and the function is addition.

Can something like this be accomplished using quantum-based methods?  This problem
was analyzed by H.-K. Lo for the case of two parties (Alice and Bob), and he showed that
in the case of two-party secure computations it cannot \cite{lo}.  In the two-party case Alice
evaluates the function, and she has one input and Bob has the other.  In one-sided secure
function evaluation only Alice learns $f(x,y)$ and in two-sided secure computation both
learn $f(x,y)$.  In both cases Alice and Bob are to learn as little about each other's input
as possible.  Lo showed that one-sided two-party quantum secure computation is, in fact,
always insecure, that there are functions for which the two-sided scenario is also insecure.

We would like to start by showing that a modification of our traveling ballot scheme will allow us to
accomplish the task described in the first paragraph of this section for a particular function,
group multiplication in the Klein 4-group,  and for participants who are curious but follow
the protocol.  This is an order four abelian group whose elements
we shall denote by $\{ e,x_{1},x_{2},x_{3}\}$.  The element $e$ is the identity, $x_{j}^{2}=e$,
and $x_{j}x_{k}=x_{l}$, where $j$,$ k,$ and $l$ are all different.  Alice, Bob and Charlie each
choose a group element, and they want to know the product of the three elements.  Donna
prepares the two-qubit state $|\Psi\rangle = (|0\rangle |1\rangle -|1\rangle |0\rangle )/ \sqrt{2}$,
keeps one qubit and sends the other to Alice.  Based on her choice of a group element, Alice
then applies an operation to the qubit using the correspondence
\begin{eqnarray}
e\rightarrow I & \hspace{1cm} & x_{1}\rightarrow \sigma_{x} \nonumber  \\
x_{2}\rightarrow \sigma_{y} & \hspace{1cm} & x_{3} \rightarrow \sigma_{z} ,
\end{eqnarray}
where $I$ is the identity, and $\sigma_{x}$, $\sigma_{y}$, and $\sigma_{z}$ are the Pauli matrices.
She then sends the qubit on to Bob, who applies an operation to the qubit based on his choice
of group element (using the same correspondence between operations and group elements),
and he then sends it on to Charlie who does the same.  Finally, Charlie
sends the particle back to Donna.  Donna measures the resulting state in the Bell basis, and
from this measurement she can determine which of the four states she has, $|\Psi\rangle$,
$(I\otimes \sigma_{x})|\Psi\rangle$, $(I\otimes \sigma_{y})|\Psi\rangle$, or
$(I\otimes \sigma_{z})|\Psi\rangle$ (each of these states is proportional to an element
of the Bell basis).  Using the
correspondence between group elements and operations she can tell what the product
of the group operations was.  For example, if she found that she had
$(I\otimes \sigma_{x})|\Psi\rangle$, then she would know the product was $x_{1}$.

This procedure is a variant of dense coding \cite{bennett}.  It is based on the fact that the
operators $\{ I,\sigma_{x}, \sigma_{y}, \sigma_{z} \}$ form a projective representation of
the Klein 4-group.  Note that during the entire procedure the reduced density matrix of
each of the participants is $\rho = I/2$, so they are able to learn nothing about what the other
participants have done.  The final state received by Donna contains no information about
who did what, it only contains information about the product of their choices of group elements.

This scheme can be generalized to any finite group and arbitrary number of participants.
Let $G$ be a group, and $g\in G\rightarrow
U(g)$, where $U(g)$ is a $D \times D$ unitary matrix, and the matrices $U(g)$ form a
$D$-dimensional representation of $G$.  Donna starts with the two-qudit state
\begin{equation}
 |\Psi\rangle = \frac{1}{\sqrt{D}} \sum_{j=0}^{D-1} |j\rangle |j\rangle .
\end{equation}
The second qudit is sent to Alice, who acts on it with $U(g_{a})$, where $g_{a}\in G$ is her
input, and then sends the qudit on to Bob.  Bob applies the operation  $U(g_{b})$, where
$g_{b}\in G$ is his input, and sends the qudit on to Charlie, who does the same, etc.
At the end of the procedure Norbert sends the qudit back to Donna who has the two-qudit
state $I\otimes U(g_{p}) |\Psi\rangle$, where
$g_{p}=g_a g_b \dots g_n$
is the product of the group elements chosen by the parties who are providing the
inputs.  A requirement is that these states are orthogonal for different group element, $g_{p}$,
so that Donna can distinguish them.   This requires that
\begin{equation}
\langle\Psi | I \otimes U(g_{2}^{-1}g_{1})|\Psi\rangle = 0
\end{equation}
for any two $g_{1}, g_{2}\in G$, such that $g_{1}\neq g_{2}$.  This condition will be fulfilled if
${\rm Tr}(U(g))=0$ for any group element not equal to the identity.  This condition is satisfied by
the regular representation of any group.   For this representation, which is, in general, reducible,
the dimension is equal to the order of the group.  In order to give an explicit description of the
matrices $U(g)$ in this representation, we order the group elements, $g_{j}$, where $j=0,\ldots
|G|-1$.  The matrix elements of $U(g_{n})$ are then given by
\begin{equation}
U(g_{n})_{jk}=\left\{ \begin{array}{cc} 1 &{\rm if}\  g_{j}^{-1}g_{k}=g_{n}  \\ 0 & {\rm otherwise}
\end{array} \right.    .
\end{equation}
It may be possible to find representations of smaller dimension that satisfy the condition,
${\rm Tr}(U(g))=0$ for any group element not equal to the identity, but we are at least assured
that if we choose the dimension equal to the order of the group, such a representation exists.

Note that if we used the regular representation in the case of the Klein 4-group, our representation
would have had dimension four, but we were able to find one that has dimension two.  The
two-dimensional representation is, in fact, a projective representation.
A projective representation
of a group is a mapping from the group to unitary matrices, $g\rightarrow U(g)$, that satisfies
\begin{equation}
U(g_{1})U(g_{2})=e^{i\omega (g_{1},g_{2})}U(g_{1}g_{2})  ,
\end{equation}
where $\omega (g_{1},g_{2})$ is a real-valued function on $G\times G$.  A projective
representation that satisfies ${\rm Tr}(U(g))=0$ for any group element not equal to the identity
will also produce states $I\otimes U(g_{p}) |\Psi\rangle$ that are mutually orthogonal.  In some
cases, the use of a projective representation will allow one to achieve this result with a smaller
dimensional space than would be possible if one restricted oneself to standard representations.

For abelian groups it is also possible to use a distributed ballot scheme.  This is because any
abelian group is isomorphic to a direct product of cyclic groups.  We distribute one particle for
each cyclic group appearing in the decomposition of the abelian group, and the parties apply
operators, similar to the voting operators in the previous section, to each particle to encode
their group element.  At the end of the procedure, all of the particles are returned to Donna,
who can then determine the product of the group elements.

Let's illustrate this procedure with a simple example.  We again consider the Klein 4-group and
the parties Alice, Bob, and Charlie, who are to choose group elements.
The Klein 4-group is isomorphic to
$Z_{2}\times Z_{2}$, whose elements can be expressed as $\{ (0,0), (0,1), (1,0), (1,1) \}$.
The following state is prepared
\begin{equation}
|\Psi\rangle = \frac{1}{2} \left( \sum_{j=0}^{1} |j\rangle^{\otimes 3} \right) \otimes
\left( \sum_{k=0}^{1} |k\rangle^{\otimes 3} \right) ,
\end{equation}
and one qubit from the first triple and one from the second is distributed to each of the three parties.  
Each party now chooses a group element and performs an operation on his or her pair of qubits
according to the correspondence
\begin{eqnarray}
(0,0) \rightarrow I\otimes I & (1,0) \rightarrow \sigma_{z}\otimes I \nonumber \\
(0,1) \rightarrow I \otimes \sigma_{z} & (1,1) \rightarrow \sigma_{z}\otimes \sigma_{z} .
\end{eqnarray}
All of the qubits are then sent to Donna.  She measures each triple in the basis
\begin{equation}
|\phi_{\pm}\rangle = \frac{1}{\sqrt{2}}(|0\rangle |0\rangle |0\rangle \pm |1\rangle |1\rangle
|1\rangle ) ,
\end{equation}
with a $|\phi_{+}\rangle$ result corresponding to a $0$ and a $|\phi_{-}\rangle$ result
corresponding to a $1$.  For example, if she obtained $|\phi_{+}\rangle$ for the first triple
and $|\phi_{-}\rangle$ for the second, this corresponds to the group element $(0,1)$.
Therefore, she is able to determine the product of the group elements chosen by Alice, Bob,
and Charlie without knowing their individual choices.

\section{Dishonest voters and eavesdroppers}
We now want to change the rules.  So far, we have been assuming that everyone was honest but
curious.  Now we want to relax that constraint.  First, we will look at the case of dishonest voters.  These
voters want to vote more than once.  We shall present a scheme that prevents them from doing that.
Another possibility is that all of the participants in the voting scheme are honest, but there is an eavesdropper
who wants to discover how one or more of the voters voted.  We shall now explore these two scenarios. 
\subsection{Dishonest voters}
One problem with the voting schemes presented so far is that there is nothing to prevent
voters from voting more than once.  If they want to vote ``yes'' more than once they simply apply
the operator corresponding to a ``yes'' vote more than once, if they want to increase the number
of``no''votes they apply the inverse of the ``yes'' operator.  One possible way of dealing with this
problem was suggested in Ref.~\cite{hillery}. In this section we will discuss variations of the
distributed-ballot and traveling-ballot that deal with this problem. 

We begin with the distributed-ballot scheme.  The ballot state is the same as in the Eq.~(\ref{ballotstate}).
In addition, the authority distributes to each voter two
voting states, which are single qudits.  The voting qudit corresponding to a ``yes'' vote is in
the state $|\psi (\theta_{y})\rangle$ and the qudit corresponding to a``no''vote is in the state
 $|\psi (\theta_{n})\rangle$, where
\begin{equation}
|\psi (\theta )\rangle = \frac{1}{\sqrt{D}}\sum_{j=0}^{D-1}e^{ij\theta}|j\rangle ,
\end{equation}
and the angles $\theta_{y}$ and $\theta_{n}$ are given by $\theta_{y}=(2\pi l_{y}/D)+\delta$
and  $\theta_{n}=(2\pi l_{n}/D)+\delta$.  The integers $l_{y}$ and $l_{n}$ and the number
$0\leq \delta < 2\pi/D$ are known only to the authority.  The voter chooses the voting particle
corresponding to his vote, and using a process much like teleportation, is able to transfer the
state of the voting qudit onto his ballot qudit.  Because Alice knows $l_{y}$, $l_{n}$,
and $\delta$ she can determine the number of ``yes'' votes.  If a voter tries to cheat and measure
the values of $\theta_{y}$ and $\theta_{n}$, he can only measure them to an accuracy of order
$2\pi /D$.  If he uses these measured values to vote, he will introduce errors.  These errors
will show up if the voting is repeated several times.  If no cheating occurred, then the result
will be the same each time.  If cheating did occur, then the results will fluctuate.
Therefore, the authority would be able to tell if someone is cheating.  In order to facilitate voting
several times, the authority can distribute several ballot states to the voters at the beginning of the
voting process and instruct them to vote the same way on each one.
Let us now examine this procedure in more detail:

\noindent
{\bf Step 0:} {\it Distribution of states} \newline
Alice distributes the entangled ballot state $|\Psi\rangle$
described in Eq.(\ref{ballotstate}) and sends to each voter two additional
qudits $|\psi (\theta_{y})\rangle$ and $|\psi (\theta_{n})\rangle$.
First, we assume that $(l_{y}-l_{n})N <D$, where, as before, $N$ is the number
of voters (Vincent.1, \dots , Vincent.N).  This condition is necessary in order that different
voting results be distinguishable. As previously mentioned, the integers $l_{y}$ and
$l_{n}$ and the angle $\delta$ are not known to the voters.

\noindent
{\bf Step 1:} {\it Voting process}\newline
Depending
on his choice the voter (Vincent.X) combines either $|\psi(\theta_y)\rangle$,
or $|\psi(\theta_n)\rangle$, with the original
ballot particle, i.e. creates a system composed from the ballot
and the voting qudits. Then he performs a two-qudit measurement of the observable
$R=\sum_{r=0}^{D-1}rP_{r}$, where
\begin{equation}
 P_r=\sum_{j=0}^{D-1}  | j+r\rangle_{b}\langle j+r|\otimes |j\rangle_{v}\langle j| ,
\end{equation}
and the subscript $b$ denotes the
ballot qudit while the subscript $v$ denotes the voting qudit.
Registering the outcome $r$ the voter applies the operation
$V_{r}=I_b\otimes \sum_{j=0}^{D-1} |j+r\rangle_v \langle j|$ to the  voting qudit.
If the voter voted ``yes'', the state of the ballot and voting state is then (up to normalization)
\begin{eqnarray}
\nonumber
V_{r}P_{r}|\Psi\rangle|\psi (\theta_{y})\rangle &=&
 \frac{1}{D}\left( \sum_{k=0}^{r-1} e^{i(D+k-r)\theta_{y}}|k\rangle^{\otimes (N+1)}\right.\\
 & &  + \left.\sum_{k=r}^{D-1} e^{i(k-r)\theta_{y}} |k\rangle^{\otimes (N+1)} \right) .
\end{eqnarray}
It is necessary to get rid of the factor $\exp (iD\theta_{y})= e^{iD\delta} $ in the first term.
After a voter has voted, he tells (publicly) the authority the value of $r$ he obtained,
because only the authority has knowledge of $\delta$ and can undo this factor.
Each voter sends both (the ballot and the voting) qudits back to the
authority.  The remaining unused qudit must be kept, or destroyed
in order to secure the privacy of the registered vote. 

\noindent
{\bf Step 2:} {\it Reading the result}\newline
When the ballot state is returned to the authority, she applies an operator
\begin{equation}
W=\prod_{k=1}^{N}W_{r_{k}}
\end{equation}
to one of the particles in the ballot state \cite{azkune}.  The integer $r_{k}$ is the value of $r$ obtained by the
$k^{\rm th}$ voter, where
\begin{equation}
W_{r}|k\rangle = \left\{ \begin{array}{cc} e^{-iD\delta}|k\rangle & 0\leq k \leq r-1 \\ |k\rangle &
r \leq k \leq D-1 \end{array} \right. \, .
\end{equation}
That removes the unwanted phase factors. The authority is then in possession of
a state consisting of $2N$ qudits. If $m_y=m$  voters voted ``yes'', $m_n=N-m$
voters voted ``no", the authority, after the application of the operator $W$, now has the state
\begin{equation}
|\Omega_{m}^{\prime}\rangle =\frac{1}{\sqrt{D}}\sum_{j=0}^{D-1}
e^{ij(m_y\theta_y+m_n\theta_n)}|j\rangle^{\otimes 2N} ,
\end{equation}
where an irrelevant global phase factor has been dropped.
The phase factor appearing in the sum can be expressed as
\begin{equation}
e^{ij(m_y\theta_y+m_n\theta_n)}=e^{ijm\Delta} e^{ijN\theta_n} ,
\end{equation}
where $\Delta=\theta_y-\theta_n = 2\pi (l_{y}-l_{n})/D$.  The factor $e^{ijN\theta_n} $ can be
removed by the authority by applying a unitary transformation that shifts $|j\rangle$ to
$e^{-ijN\theta_n} |j\rangle$ to one of the qudits.
This finally leaves the authority with the state
\begin{equation}
\label{omegaq}
|\Omega_{q}\rangle =\frac{1}{\sqrt{D}}\sum_{j=0}^{D-1} e^{2\pi ijq/D}|j\rangle^{\otimes 2N} ,
\end{equation}
where $q=m(l_{y}-l_{n})$.  These states are orthogonal
for different values of $q$, for $q$ an integer between $0$ and
$D-1$ (we need to choose $| l_{y}-l_{n}|$ and $D$ to guarantee that $q$ is in this range),
so we see that from the state $|\Omega_q\rangle$ the authority can
determine the value of $q$ corresponding to this state.  This allows
her to determine $m$, because she knows both $l_{y}$ and $l_{n}$.  Note that $q$ should
always be a multiple of $l_{y}-l_{n}$ if the voters are using their proper ballot states.
If after measuring the ballot state, the authority finds a value of $q$
that is not a multiple of $l_{y}-l_{n}$, then she knows that someone has cheated.
Let us note that the total measurement is described by projective operations
$M_q=|\Omega_q\rangle\langle\Omega_q|$, for $0\leq q \leq D-1$ and a multiple of $l_{y}-l_{n}$, 
and $M_{\rm error}=I-\sum_q M_q$.

A similar procedure works for a traveling-ballot scheme.  In this case, the previous traveling-ballot scheme
is modified so that votes are recored by means of
a rotation rather than as a shift.  We start with the ballot state in Eq.\ (\ref{ballot_travel}),  and as in the
distributed-scheme voting particles in the states $|\psi (\theta_{y})\rangle$ and $|\psi (\theta_{n})\rangle$
are distributed to the voters.  We still have $\theta_{y}=(2\pi l_{y}/D)+\delta$ and  
$\theta_{n}=(2\pi l_{n}/D)+\delta$.   A voter now combines the ballot state with the voting particle representing
his choice and measures $R$ as before.  Suppose he wants to vote ``yes'' and the result $r$ is obtained upon
measuring $R$.  The state is then
\begin{eqnarray}
P_{r}|\Psi\rangle|\psi (\theta_{y})\rangle &=&
 \frac{1}{D}\left( \sum_{j=0}^{r-1} e^{i(D+j-r)\theta_{y}}|j\rangle_{a}|j\rangle_{b}|j-r+D\rangle_{v}\right. \nonumber \\
 & &  + \left.\sum_{j=r}^{D-1} e^{i(j-r)\theta_{y}} |j\rangle_{a}|j\rangle_{b}|j-r\rangle_{v} \right) .
\end{eqnarray}
The voter now tells the authority the value of $r$, and the authority applies the operator $W_{r}$ to the particle
in her possession.  This removes the unwanted factor of $e^{iD\theta_{y}}$ in the first term.  The voter now
applies the operator $U_{r}$ to the ballot and voting particle, where
\begin{equation}
U_{r}|j\rangle_{b}|j+r-D\rangle_{v}  =  |j\rangle_{b}|0\rangle_{v}   ,
\end{equation}
for $0\leq j \leq r-1$, and
\begin{equation}
U_{r}|j\rangle_{b}|j-r\rangle_{v}  =  |j\rangle_{b}|0\rangle_{v} ,
\end{equation}
for $r\leq j \leq D-1$.  This has the effect of disentangling the voting particle from the rest of the state,
\begin{eqnarray}
U_{r}W_{r}P_{r}|\Psi\rangle|\psi (\theta_{y})\rangle &= 
 \frac{1}{D}\left( \sum_{j=0}^{r-1} e^{i(j-r)\theta_{y}}|j\rangle_{a}|j\rangle_{b} \right. \nonumber \\
 &   + \left.\sum_{j=r}^{D-1} e^{i(j-r)\theta_{y}} |j\rangle_{a}|j\rangle_{b} \right)  |0\rangle_{v}  .
\end{eqnarray}
The ballot particle is now passed on to the next voter, who repeats the procedure.  At the end of the voting,
the ballot particle is returned to the authority, who then has the state 
\begin{equation}
|\Omega_{m}^{\prime\prime}\rangle =\frac{1}{\sqrt{D}}\sum_{j=0}^{D-1}
e^{ij(m_y\theta_y+m_n\theta_n)}|j\rangle_{a}|j\rangle_{b} ,
\end{equation}
up to a global phase factor.  From there on the analysis is the same as in the distributed ballot case.

As we discussed at the beginning of this subsection a voter who wants to vote more than once is
faced with the problem of determining what $\theta_{y}$ or $\theta_{n}$ are, and this cannot
be done from just a single state. However, there is a small chance that the
cheating won't be detected and therefore, the voting has to be performed several times.
However, just a single difference in outcomes means that someone is cheating. The details
of an attack by a cheater and its consequences are described in Appendix.

\subsection{Eavesdropper}
Now we shall consider an attack by an external eavesdropper, who wants to learn how one
of the participants voted.  The actual participants in the protocol are assumed to be honest but curious.

First, let us consider the traveling-ballot scheme.
Suppose  an eavesdropper, Eve wants to know how the second voter, Vincent.2, voted.
She intercepts the ballot qudit just before it is due to be received
by Vincent.2 and sends it on to Vincent.3.  To Vincent.2 she sends her own qudit, which is in the
state $|0\rangle$.  After Vincent.2 votes, she intercepts the qudit and measures it; if it is in the state
$|0\rangle$, Vincent.2 voted ``no'', if it is in the state $|1\rangle$, then Vincent.2 voted ``yes''.  This
type of attack seems to be very hard to prevent.  One possibility, which is very expensive in terms
or resources, is to use teleportation.  If successive voters share entangled
two qudit states of the form given in Eq.\ (\ref{ballot_travel}), they can then teleport the ballot state to each other rather than physically send the ballot particle.  This procedure would prevent the 
of man-in-the-middle attack just described, but requires that the participants originally shared many qudit pairs and
used entanglement purification to bring any correlations with outside systems, such as those
possessed by an eavesdropper, to acceptable levels.  Therefore, this approach is not a particularly 
desirable one.

A distributed-ballot scheme seems to offer more possibilities.  In order to illustrate this, we shall compare
the vulnerability of a classical and a quantum scheme to eavesdropping.  We shall consider the case 
in which there is an eavesdropper, Eve, who wants to find out how Vincent.1 voted. 

Our classical scheme is a variant of one proposed in the paper by Dolev {\it et al.} \cite{dolev}.
There are two authorities, one who generates ballots and one who counts the votes, and there
are $N$ voters.  The first authority generates $N$ ballots, one for each voter, and on each ballot
an integer between $0$ and $N$ is written.  These numbers have the property that their sum is
equal to zero modulo $N+1$.  When each voter receives his ballot, he does
nothing to vote ``no'' and adds $1$ to vote ``yes.''  The ballots are all sent to the second
authority, who simply adds all of the numbers modulo $N+1$, with the result being the number
of ``yes'' votes.

The second authority does not know how any of the individual voters voted, because she
does not know the original integers written on the ballots.  If fact, she has no information
about how the voters voted, if each set of ballots (that is, each sequence of $N$
integers whose sum is zero modulo $N+1$) is equally likely.  We can see this as follows.
We can represent the initial state of the ballots by a sequence of $N$ integers, each of
which is between $0$ and $N$ and whose sum is zero modulo $N+1$.  Similarly, we can
represent the final state (after voting) of the ballots by a sequence of $N$ integers, each of
which is between $0$ and $N$ and whose sum is $m$ modulo $N+1$, where $m$ is the
number of ``yes'' votes.  The set of voters who voted ``yes'' can be represented by a sequence
of ones and zeroes, ones denoting the voters who voted ``yes,'' of length $N$.  Now for
each sequence of $N$ integers whose sum is equal to $m$ mod $N+1$, and each sequence
of length $N$ consisting of $m$ ones and $N-m$ zeroes, there is a sequence of $N$ integers
whose sum is $0$ mod $N+1$( found simply by subtracting the second sequence from the first).
Thus with no knowledge of the initial ballot set, all we can conclude from a final ballot set
whose numbers sum to $m$ mod $N+1$,  is that some subset consisting of $m$ voters
voted ``yes.''   Therefore, the voting information, that is,  who voted how, is protected from
the curiosity of the authorities.

Now let us add the eavesdropper.  Eve wants to know how voter number $1$ (Vincent 1) voted, and she
has an excellent method of doing so.  She intercepts the ballot going to Vincent 1,
records the number on it, and sends the ballot on to him.  Vincent 1 votes, and Eve again
intercepts the ballot, notes the result, and sends it on to the second authority.  Eve now knows
how Vincent 1 voted, and her intervention has not been detected.

Next let us consider the quantum scheme.  The ballot state is the $N$ qudit state given by Eq.\
(\ref{ballotstate}).  We shall assume that the same authority prepares the
ballot state and later measures it to count the votes.  As before, a qudit from the ballot
state is sent to each voter, and if they wish to vote ``no,'' they do nothing, and if they
wish to vote ``yes,'' they apply the operator $F$.  

Let us now suppose that Eve wants to determine how Vincent 1 voted and not be detected.  One 
way of doing this is the following.  Eve intercepts ballot particle $1$ on its way to Vincent 1 and
entangles it with an ancilla.  In particular, suppose the ancilla is a qudit initially in the state 
\begin{equation}
|\psi\rangle_{E} = \frac{1}{\sqrt{D}} \sum_{k=0}^{D-1} |k\rangle_{E} ,
\end{equation}
and the entangling operation is the swap operator, $U_{swap}|k\rangle_{E}\otimes |j\rangle_{1}
= |j\rangle_{E} \otimes |k\rangle_{1}$.  After this is done, the ballot plus ancilla state is
\begin{equation}
|\Psi^{\prime}\rangle = \frac{1}{D} \sum_{k=0}^{D-1} \sum_{j=0}^{D-1} |j\rangle_{E}|k\rangle_{1}
|j\rangle^{\otimes (N-1)}  .
\end{equation}
After the voting, the state becomes 
\begin{equation}
|\Psi^{\prime\prime}\rangle = \frac{1}{D} \sum_{k=0}^{D-1} \sum_{j=0}^{D-1} e^{2\pi i m_{1}k/D}
e^{2\pi i mj/D} |j\rangle_{E}|k\rangle_{1}|j\rangle^{\otimes (N-1)}  ,
\end{equation}
where $m_{1}=0,1$ is the vote of Vincent 1 and $m$ is the sum of the rest of the votes.  Now, Eve
again intercepts ballot particle number $1$ on its way to the authority and again applies the swap
operator to particle $1$ and the ancilla. The state of the system is now
\begin{equation}
|\Psi^{\prime\prime\prime}\rangle = \frac{1}{D}\left( \sum_{k=0}^{D-1} e^{2\pi i m_{1}k/D}  |k \rangle_{E}
\right) \sum_{j=0}^{D-1}  e^{2\pi i mj/D} |j\rangle_{1} |j\rangle^{\otimes (N-1)}  .
\end{equation}
Now Eve can measure the ancilla particle to determine $m_{1}$.  Once she has done so, she applies
the appropriate operator to particle $1$, nothing if $m_{1}=0$ and $F$ if $m_{1}=1$, and sends the
particle to the authority.  At this point she knows how Vincent 1 voted, and her presence has not
been detected.  

So far, the quantum scheme seems just as vulnerable as the classical one.  We can defend against
the kind of attack discussed above by adding an additional element.  Before the voting occurs, the
voters are divided up into pairs.  Who is in which pair is not public knowledge.  This can be accomplished
if the authority and voters share a secure key.  This would allow the authority to tell each voter with whom 
they are paired in a secure fashion.  The voters in each pair must come together, perhaps at a polling place,
where they can perform a joint measurement on their ballot particles.  If
there has been no tampering, these measurements do not change the state of the system, and the 
voting proceeds as usual.  If the measurements detect tampering the procedure is aborted.  One could
group the voters into larger sets and perform correspondingly larger collective measurements.
Pairs minimizes the complexity of the collective measurements, and it means that each voter has
to meet with only one other voter to perform the collective measurement.  It is important that Eve
not know which voters have been assigned to the pairs.  If she did, she could perform an attack
using swap operators on a pair, which is very similar to the attack discussed above, and learn how
the pair voted.  Using this attack, she would, however, not learn how an individual voted.

Having the some of the voters come together is awkward, but for the type of check we are discussing it
seems to be necessary.  The basic idea is that if Eve wants to determine how an individual, or set of individuals,
voted, she has to break the symmetry of the ballot state.  In order to detect this, the voters have to test the
symmetry of the ballot state, and this seems to require a collective measurement.  An alternative would be to use
teleportation, and have one member of a pair teleport the state of the ballot particle to the other member of the pair,
who could then perform the collective measurement and teleport the particle back if the measurement was
successful.  This would be, however, very expensive in terms of the number of entangled pairs required.

In order to examine this type or eavesdropping attack in detail, let us suppose that one of the pairs consists of voters $1$ and $2$.  When they receive their ballot particles, they perform the measurement corresponding to the 
projection operator
\begin{equation}
P_{12}=\sum_{j=0}^{D-1} |j\rangle_{1}\langle j| \otimes |j\rangle_{2}\langle j| .
\end{equation}
If they get $1$, they proceed to voting, if they get $0$, they abort the procedure.  

Now suppose that an eavesdropper, who wants to find out how voter $1$ votes, has intercepted
the ballot particle destined for voter $1$ and entangled it with an ancilla in her possession, and 
then sent the ballot state on to voter $1$.  We assume that the entanglement has been accomplished
by means of a unitary operator $U_{E1}(|0\rangle_{E}\otimes |j\rangle_{1}) = |\phi_{j}\rangle_{E1}$.
The state of the $N$ voters plus the ancilla is now
\begin{equation}
|\Psi^{\prime}\rangle = \frac{1}{\sqrt{D}} \sum_{j=0}^{D-1} |\phi_{j}\rangle_{E1}\otimes 
|j\rangle^{\otimes (N-1)} .
\end{equation}
Eve's plan is to measure the ancilla after the voting has occurred to gain information about how
voter $1$ voted.  The probability of not detecting the eavesdropping is just 
$\langle\Psi^{\prime}|P_{12}|\Psi^{\prime}\rangle$, which can be expressed as
\begin{equation}
 \langle\Psi^{\prime}|P_{12}|\Psi^{\prime}\rangle = \frac{1}{D}\sum_{j=0}^{D-1}
 \,_{E1}\langle \phi_{j}| (I_{E}\otimes |j\rangle_{1}\langle j|) |\phi_{j}\rangle_{E1} ,
\end{equation}
where $I_{E}$ is the identity on the ancilla space.  Eve would like this quantity to be equal to $1$,
i.e.\ she does not want to be detected.  For that to be true, we must have $ |\phi_{j}\rangle_{E1}
= |\eta_{j}\rangle_{E}\otimes |j\rangle_{1}$ for some ancilla states $|\eta_{j}\rangle$.  If this is the case, 
the state after the voting has taken place is 
\begin{equation}
|\Psi^{\prime\prime}\rangle = \frac{1}{\sqrt{D}} \sum_{j=0}^{D-1} e^{2\pi i mj/D}
|\eta_{j}\rangle_{E}\otimes |j\rangle^{\otimes N} ,
\end{equation}
if $m$ voters voted ``yes.''  Tracing out all of the voters except for voter $1$, we find that the density
matrix for voter $1$ and the ancilla state is
\begin{equation}
\label{rhoE1}
\rho_{E1} = \frac{1}{D}\sum_{j=0}^{D-1} |\eta_{j}\rangle_{E}\langle \eta_{j}| \otimes |j\rangle_{1}\langle j| ,
\end{equation}
which contains no information about the votes.  That means that even if Eve intercepts ballot particle
$1$ after the vote and performs an entangling operation on it and the ancilla, she will learn nothing
about the vote.  So, if the eavesdropper is undetectable, she gains
no information about the voting, and if she gains information about the voting, she can be detected.

Note that even in the general case when $|\phi_{j}\rangle_{E1}$ is not a product state, if voters
$1$ and $2$ obtained one when they measured $P_{12}$, then it will be after the measurement.
This follows from the fact that $(I_{E}\otimes |j\rangle_{1}\langle j|) |\phi_{j}\rangle_{E1} = 
|\mu_{j}\rangle_{E} \otimes |j\rangle_{1}$, where $|\mu_{j}\rangle_{E}$ is an unnormalized ancilla
state.  Then the density matrix for the ancilla and particle $1$ will look the same as in 
Eq.\ (\ref{rhoE1}) except that $|\eta_{j}\rangle$ will be replaced by $|\mu_{j}\rangle_{E}$, and
there will be an overall normalization factor.  It still does not contain any information on the
voting.

Finally, let us find the probability of Eve being detected in the scheme that made use of the swap
operator.  A short calculation shows that the probability of Eve not being detected is 
\begin{equation}
\langle\Psi^{\prime}|P_{12}|\Psi^{\prime}\rangle = \frac{1}{D} ,
\end{equation}
so that the probability of her being detected is $1-(1/D)$.  Therefore, it is quite likely that this type 
of tampering by Eve will be detected.

\section{Conclusion}
We have shown how quantum mechanics can be of use in maintaining privacy in tasks such as
anonymous voting and in a special case of multi-party function evaluation. The voting scheme
we described was introduced in Ref.~\cite{hillery}. In this paper we provide a more detailed
discussion of that scheme, including a more detailed analysis of some aspects of its security.

The schemes presented here need much more examination in order to determine how secure they
are under different kinds of cheating and eavesdropping attacks.  It could be the case that
quantum resources themselves will not provide us with any additional feature that will result
in more effective and secure anonymous voting protocol. However, it is useful to
think about how quantum resources can be applied to such complex problem.  Such efforts
can bring results that can potentially enhance privacy in less
complex cryptographic tasks. We hope that what has been presented here will provide a framework for 
thinking about these issues.

\noindent\newline{\bf Acknowledgement}\\
We thank G.~Azkune and J.~Vaccaro for helpful comments.
This work was supported in part by the
European Union projects  CONQUEST and QAP,  by
the Slovak Academy of Sciences via the project CE-PI/2/2005, by the
APVT  via the project QIAM and by the National Science Foundation under grant number PHY-0903660.

\appendix
\section{Cheating by multiple votes}
Let us look at cheating in more detail.  We shall assume that one of the voters, whom we shall call
Vincent.X, is dishonest, and that he wants to vote ``no'', and in addition he wants to replace ``yes''
votes with ``no'' votes.  He employs a measurement to determine $\theta_{y}$ and $\theta_{n}$,
which is described by the POVM operators $E(\theta )= (D/2\pi )|\Phi (\theta )\rangle
\langle\Phi (\theta )|$, where
\begin{eqnarray}
\nonumber
|\Phi (\theta )\rangle = \frac{1}{\sqrt{D}} \sum_{j=0}^{D-1}e^{ij\theta}|j\rangle  .
\end{eqnarray}
This is a phase estimation measurement, and the probability distribution for the measurement
result $\theta$ in the state $|\psi\rangle$ is $p(\theta )=\langle\psi |E(\theta )|\psi\rangle$.  If Vincent.X
obtains the values $\theta_{y}^{\prime}$ and $\theta_{n}^{\prime}$ from his measurements of the
voting particles, the probability distributions for these results are
\begin{eqnarray}
p_{y}(\theta_{y}^{\prime}) & = & \langle\psi (\theta_{y})|E(\theta_{y}^{\prime})|\psi (\theta_{y})\rangle  = \frac{1}{2\pi D}
\left| \sum_{j=0}^{D-1}e^{ij(\theta_{y}^{\prime}-\theta_{y})} \right|^{2} ;\nonumber \\
p_{n}(\theta_{n}^{\prime}) & = & \langle\psi (\theta_{n})|E(\theta_{n}^{\prime})|\psi (\theta_{n})\rangle  = \frac{1}{2\pi D}
\left| \sum_{j=0}^{D-1}e^{ij(\theta_{n}^{\prime}-\theta_{n})} \right|^{2}  .
\nonumber
\end{eqnarray}
Note that these functions are peaked about the values $\theta_{y}$ and $\theta_{n}$, respectively.
In order to vote``no'' Vincent.X prepares a particle in the state $|\psi(\theta_{n}^{\prime})\rangle$ and
carries out the usual voting procedure with it.  He then applies the operator
\begin{eqnarray}
U(\theta_{y}^{\prime},\theta_{n}^{\prime})=\sum_{k=0}^{D-1}
e^{ik(\theta_{n}^{\prime}-\theta_{y}^{\prime})} |k\rangle\langle k| ,
\nonumber
\end{eqnarray}
to his ballot qudit $s$ times, which has the effect of removing $s$ ``yes'' votes and adding $s$``no''votes.
He then sends his ballot and voting qudits back to the authority.

We want to see how Vincent.X's cheating affects the measurement the authority makes to determine
the number of ``yes'' votes.  We shall assume, for the sake of simplicity, that Vincent.X is the last person
to vote, and that $m_{y}$ of the previous voters voted ``yes'', and $m_{n}$ voted no, where
$m_{y}+m_{n}=N-1$.  The order in which the voters vote makes no difference to the final result, so this 
assumption is made for the sake of notational convenience.
In addition, we shall also assume that the other voters have reported their results from measuring the observable $R$,
and that the necessary corrections have been applied.
This means that the state of the ballot and voting particles just before
it reaches Vincent.X is
\begin{eqnarray}
|\Xi_{1}\rangle = \frac{1}{\sqrt{D}}\sum_{j=0}^{D-1}e^{ij(m_{y}\theta_{y}+m_{n}\theta_{n})}
|j\rangle^{\otimes (2N-1)} .
\nonumber
\end{eqnarray}
As stated in the previous paragraph, Vincent.X now prepares a qudit in the state $|\psi(\theta_{n}^{\prime})\rangle$
and applies the usual voting procedure.  Let us suppose
that when he measures the observable $R$ he obtains the value $r$.  After he applies
$U(\theta_{y}^{\prime},\theta_{n}^{\prime})$ $s$ times the state of the ballot and voting
particles is

\begin{widetext}
\begin{eqnarray}
|\Xi_{2}\rangle =  \frac{1}{\sqrt{D}}\sum_{j=0}^{r-1}e^{i(D+j-r)\theta_{n}^{\prime}}
e^{isj(\theta_{n}^{\prime}-\theta_{y}^{\prime})} e^{ij(m_{y}\theta_{y}+m_{n}\theta_{n})}
|j\rangle^{\otimes 2N}  
  +\frac{1}{\sqrt{D}}\sum_{j=r}^{D-1}e^{i(j-r)\theta_{n}^{\prime}}
e^{isj(\theta_{n}^{\prime}-\theta_{y}^{\prime})} e^{ij(m_{y}\theta_{y}+m_{n}\theta_{n})}
|j\rangle^{\otimes 2N} .
\nonumber
\end{eqnarray}
This is the state possessed by the authority (Alice) after the ballot and voting particles have been returned
to her.  Alice now uses Vincent.X's measurement result, $r$, to correct the state.  After having
done so, and after removing unimportant phase factors, the authority has the state
\begin{eqnarray}
|\Xi_{3}\rangle  =  \frac{1}{\sqrt{D}}e^{iD(\theta_{n}^{\prime}-\delta )}\sum_{j=0}^{r-1}
e^{ij[s(\theta_{n}^{\prime}-\theta_{y}^{\prime})+m\Delta + \theta_{n}^{\prime}-\theta_{n}]}
|j\rangle^{\otimes 2N} 
 + \frac{1}{\sqrt{D}} \sum_{j=r}^{D-1}
e^{ij[s(\theta_{n}^{\prime}-\theta_{y}^{\prime})+m\Delta + \theta_{n}^{\prime}-\theta_{n}]}
|j\rangle^{\otimes 2N} ,
\nonumber
\end{eqnarray}
where $\Delta =\theta_{y}-\theta_{n}$ and we have set $m=m_{y}$.

Alice now measures the state $|\Xi_{3}\rangle$ in the $|\Omega_{q}\rangle$ basis
[see Eq.\ (\ref{omegaq})] in order to determine the number of ``yes'' votes.  If there were no
cheating she would find $q=m$ with certainty.  With cheating, however, this is no longer the
case, and this is what tells Alice that cheating has taken place.  The voting is repeated
several times, and if she finds different values of $q$, then she knows cheating has taken place.
In order to find the probability distribution for $q$ assuming that Vincent.X measured $r$ for the
observable $R$ and that $m$ people voted ``yes'', which we shall denote by $p(q|r,m))$, we first
note that the probability that the authority finds the value $q$ given that Vincent.X measured the
values $\theta_{y}^{\prime}$, $\theta_{n}^{\prime}$, and $r$, and that $m$ people voted ``yes''
is given by
\begin{eqnarray}
\nonumber
p(q|r,m,\theta_{y}^{\prime},\theta_{n}^{\prime})  =  |\langle\Omega_{q}|\Xi_{3}\rangle |^{2}
 & = & \frac{1}{D^{2}}\left| e^{iD(\theta_{n}^{\prime}-\delta )}\sum_{j=0}^{r-1}
e^{ij[s(\theta_{n}^{\prime}-\theta_{y}^{\prime})+\theta_{n}^{\prime}+\phi ]} \right. 
  + \left. \sum_{j=r}^{D-1}
e^{ij[s(\theta_{n}^{\prime}-\theta_{y}^{\prime})+\theta_{n}^{\prime}+\phi ]} \right|^{2} ,
\nonumber
\end{eqnarray}
where $\phi = m\Delta -\theta_{n} -(2\pi q/D)$.  We then have that
\begin{eqnarray}
p(q|r,m) = \int_{0}^{2\pi}d\theta_{y}^{\prime} \int_{0}^{2\pi}d\theta_{n}^{\prime}
p(q|r,m,\theta_{y}^{\prime},\theta_{n}^{\prime}) p_{y}(\theta_{y}^{\prime}) p_{n}(\theta_{n}^{\prime}) .
\nonumber
\end{eqnarray}

Let us consider a particular case in which $D=N+1$, $l_{y}=1$, $l_{n}=0$, and $s>D/2$.  This choice
of $s$ is one that Vincent.X might make if he thought that that there will be a majority of
``yes'' votes, and he wants to make sure that the measure being voted upon loses.  One then
finds that
\begin{eqnarray}
p(q|r,m) = \frac{1}{D} \left\{ 1+\frac{2(D-s)[(D-2)(D-s-1)+(s+1)]}{D^{3}}
\cos [2\pi (m-s-q)/D] \right\} .
\nonumber
\end{eqnarray}
\end{widetext}
Note that while this distribution has a maximum at $q=m-s$, which is the result Vincent.X desires, it is
very broad.  That means that when the authority measures $q$ several times she will find a spread
of values, showing her that someone is cheating.



\begin{thebibliography}{99}
\bibitem{gisin}
N.~Gisin, G.~Ribordy, W.~Tittel, and H.~Zbinden,
{\it Quantum cryptography}, Rev.\ Mod.\ Phys.\ {\bf 74}, 145 (2002).
\bibitem{chaum}
D.~Chaum,
{\it The dinning cryptographers problem: unconditional sender and recipient untraceability},
Journal of Cryptology {\bf 1}, 65 (1988).
\bibitem{bos}
J.N.E.~Bos, {\it Practical Privacy},
(PhD thesis, Eindhoven University of Technology, ISBN 9061964059, 1992)
\bibitem{christandl}
M.~Christandl and S.~Wehner,
{\it Quantum anonymous transmission}, Proc. of 11th ASIACRYPT, Springer Lecture notes in Computer
Science 3788, 217 (2005) and quant-ph/0409201 (2004).
\bibitem{bouda}
J.~Bouda and J.~\v{S}projcar,
{\it Anonymous transition of quantum information} in Proceedings of the First International Conference on 
Quantum, Nano and Micro Technologies (IEEE Computer Society Press 2007) P. 12, and 
quant-ph/0512122 (2005).
\bibitem{brassard}
{\it Anonymous quantum communication}
G.\ Brassard, A.\ Broadbent, J.\ Fitzsimmons, S.\ Gambs, and A.\ Tapp, Proc. of 13th ASIACRYPT, Springer
 Lecture notes in Computer Science 4833, 460 (2007) and quant-ph/0706.2356 (2007).
\bibitem{schneier}
B.~Schneier,
{\it Applied Cryptography}
(Wiley, New York, 1996), page125.
\bibitem{vaccaro}
J.A.~Vaccaro, J.~Spring, and A.~Chefles,
{\it Quantum protocols for anonymous voting and surveying},
Phys. Rev. A {\bf 75}, 012333 (2007).
\bibitem{hillery}
M.~Hillery, M.~Ziman, V.~Bu\v{z}ek, and M.~Bielikov{\'a},
{\it Towards quantum-based privacy and voting},
Phys.\ Lett.\ A {\bf 349}, 75 (2006) and quant-ph/0505041 (2005).
\bibitem{dolev}
S.~Dolev, I.~Pitowski, and B.~Tamir,
{\it A quantum secret ballot},
quant-ph/0602087 (2006).
\bibitem{helstrom}
C.\ W.\ Helstrom, \emph{Quantum Detection and Estimation Theory} (Academic Press,
New York, 1976).
\bibitem{bergou}
For a review of state discrimination see
Janos Bergou, Ulrike Herzog, and Mark Hillery in \emph{Quantum State Estimation} edited
by m.\ G.\ A. Paris and J.\ \v{R}eha\v{c}ek (Springer Heidelberg 2004), p. 417.
\bibitem{lo}
H.-K.~Lo,
{\it Insecurity of quantum secure computations},
Phys.\ Rev.\ A {\bf 56}, 1154 (1997).
\bibitem{bennett}
C.H.~Bennett and S.J.~Wiesner,
{\it Communication via one- and two-particle operators on Einstein-Podolsky-Rosen states},
Phys.\ Rev.\ Lett.\ {\bf 69}, 2881 (1992).
\bibitem{azkune}
G.~Azkune and J.~Vaccaro, private communication. This step fixes an error in reference 9.
\end{thebibliography}
\end{document}